\documentclass[12pt]{article}
\usepackage{pqft}
\usepackage{graphicx}

\def\lsim{\mathrel{\rlap{
\lower4pt\hbox{\hskip-3pt$\sim$}}
    \raise1pt\hbox{$<$}}}     
\def\gsim{\mathrel{\rlap{
\lower4pt\hbox{\hskip-3pt$\sim$}}
    \raise1pt\hbox{$>$}}}     
\def\scr#1{\mbox{\scriptsize #1}}

\begin{document}

\title{Three-Fluid  Simulations of
Relativistic Heavy-Ion Collisions}
\author{V.D.~Toneev\inst{1,2}, Yu.B.~Ivanov\inst{2,3},
E.G.~Nikonov\inst{1,2}, W.~N\"orenberg\inst{2},
V.N.~Russkikh\inst{2,3}}
\institutes{\inst{1}Joint Institute for Nuclear Research,
 141980 Dubna, Moscow Region, Russia, \inst{2}Gesellschaft
 f\"ur Schwerionenforschung mbH, Planckstr.$\!$ 1,
64291 Darmstadt, Germany \inst{3}Kurchatov Institute, Kurchatov
sq.$\!$ 1, Moscow 123182, Russia} \maketitle \abstract{ A
relativistic 3-fluid 3D hydrodynamic  model has been developed for
describing heavy-ion collisions at incident energies between few
and $\sim$ 200 A$\cdot$GeV. In addition to two baryon-rich fluids
which simulate mutually decelerating counterflows of target and
projectile nucleons, the new model incorporates evolution of a
third, retarded  baryon-free fluid  created by this decelerated
baryonic matter. Different equations of state, including those
with the deconfinement phase transition, are treated. A reasonable
agreement with experiment is illustrated  by proton rapidity
spectra, their dependence on  collision centrality and beam
energy.}

\authorrunning{V.D.~Toneev, Yu.B.~Ivanov, E.G.~Nikonov, W.~N\"orenberg,
V.N.~Russkikh}
\titlerunning{Three-Fluid Simulation of
 Relativistic Heavy-Ion Collisions}

\section{Introduction}
During past twenty years, hydrodynamics proved to be quite a
reasonable tool for describing heavy-ion collisions at moderate
energies (see for example references in \cite{RIN03}). From
general point of view this application of hydrodynamics seemed to
be more successful at higher energies when the number of produced
 particles gets  larger. However, as  it has been noted
by D.I.~Blokhintsev~\cite{Bl57}  in the beginning of the
multiple-production era, the uncertainty principle introduces
certain restrictions to this extension. Though a general
conclusion of Ref.~\cite{Bl57}, that the very beginning of the
expansion stage should be considered quantum--mechanically, is
valid till now, the modern quark--gluon picture of interactions
shifts the results of Blokhintsev's simplified estimate to far
ultra-relativistic energies. On other hand, it became clear that
non-equilibrium processes are very important at this stage and
that only a part of the total collision energy can be thermalized.
In this respect the standard one-fluid description assuming
instantaneous local equilibrium seems to be quite limited.

In the present paper  we demonstrate first results of a
relativistic 3-fluid 3-dimensional hydrodynamic code developed for
describing highly relativistic nucleus--nucleus collisions, i.e.
up to energies reached at the CERN SPS where a partial
deconfinement of hadrons may occur.  Below, basic features of the
hydrodynamic model will be presented and a numerical solution of
3-fluid hydrodynamics will first be compared with observables.

\section{3-Fluid Hydrodynamic Model}

A specific feature of the dynamic 3-fluid description is a finite
stopping  power resulting in a counter-streaming regime of leading
baryon-rich matter. This counter-streaming behavior is supported
by experimental rapidity distributions in nucleus--nucleus
collisions and  simulated by introducing the multi--fluid concept.
The basic idea of a 3-fluid approximation to heavy-ion collisions
\cite{MRS88,I87} is that at each space-time point $x=(t,{\bf x})$
the distribution function of baryon-rich  matter,
$f_{\scr{b}}(x,p)$, can be represented by a sum of two distinct
contributions
\begin{eqnarray}
\label{t1}
f_{\scr{b}}(x,p)=f_{\scr p}(x,p)+f_{\scr t}(x,p),
\end{eqnarray}
initially associated with constituent nucleons of the projectile
(p) and target (t) nuclei. In addition, newly produced particles,
populating the mid-rapidity region, are associated with a fireball
(f) fluid described by the  distribution function $f_{\scr
f}(x,p)$. Note that both the baryon-rich and fireball fluids may
consist of any type of hadrons  and/or partons (quarks and
gluons), rather then only nucleons and pions. However, here and
below we suppress the species label at the distribution functions
for the sake of transparency of the equations.

Using the standard procedure for deriving hydrodynamic equations
from the coupled set of relativistic Boltzmann equations with the
above-introduced distribution functions $f_\alpha$ ($\alpha=$p, t,
f), we arrive at equations for the baryon charge conservation
   \begin{eqnarray}
   \label{eq8}
   \partial_{\mu} J_{\alpha}^{\mu} (x) &=& 0,
   \end{eqnarray}
( for $\alpha=$p and t) and the energy--momentum  conservation of
the fluids
   \begin{eqnarray}
   \partial_{\mu} T^{\mu\nu}_{\scr p} (x) &=&
-F_{\scr p}^\nu (x) + F_{\scr{fp}}^\nu (x),
   \label{eq8p}
\\
   \partial_{\mu} T^{\mu\nu}_{\scr t} (x) &=&
-F_{\scr t}^\nu (x) + F_{\scr{ft}}^\nu (x),
   \label{eq8t}
\\
   \partial_{\mu} T^{\mu\nu}_{\scr f} (x) &=&
F_{\scr p}^\nu (x) + F_{\scr t}^\nu (x) - F_{\scr{fp}}^\nu (x) -
F_{\scr{ft}}^\nu (x).
   \label{eq8f}
   \end{eqnarray}
Here $J_{\alpha}^{\mu}=n_{\alpha}u_{\alpha}^{\mu}$ is the baryon
current defined in terms of  proper baryon density $n_{\alpha}$
and hydrodynamic 4-velocity $u_{\alpha}^{\mu}$ normalized as
$u_{\alpha\mu}u_{\alpha}^{\mu}=1$. Eq.~(\ref{eq8}) implies that
there is no baryon-charge exchange between p- and t-fluids, as
well as that the baryon current of the fireball fluid is
identically zero, $J_{\scr f}^{\mu}=0$.  The energy--momentum
tensors $T^{\mu\nu}_\alpha$ in Eqs.~(\ref{eq8p})--(\ref{eq8t}) are
affected by friction forces $F_{\scr p}^\nu$, $F_{\scr t}^\nu$,
$F_{\scr{fp}}^\nu$ and $F_{\scr{ft}}^\nu$. Friction forces between
baryon-rich fluids, $F_{\scr p}^\nu$ and $F_{\scr t}^\nu$,
partially transform the collective incident energy into thermal
excitation of these fluids ($|F_{\scr p}^\nu - F_{\scr t}^\nu|$)
and give rise to particle production into the fireball fluid
$(F_{\scr p}^\nu + F_{\scr t}^\nu)$, see Eq.~(\ref{eq8f}).
$F_{\scr{fp}}^\nu$ and $F_{\scr{ft}}^\nu$ are associated with
friction of the fireball fluid with the p- and t-fluids,
respectively. Note that Eqs.~(\ref{eq8p})--(\ref{eq8f}) satisfy
the  total energy--momentum conservation
\begin{eqnarray}
\partial_\mu (T^{\mu\nu}_{\scr p} +
T^{\mu\nu}_{\scr t} + T^{\mu\nu}_{\scr f}) = 0.
\label{eq10}
\end{eqnarray}

In terms of the proper energy density, $\varepsilon_\alpha$, and
pressure, $P_\alpha$, the energy--momentum tensors of the
baryon-rich fluids ($\alpha=$p and t) take the conventional
hydrodynamic form
   \begin{eqnarray}
T^{\mu\nu}_\alpha= (\varepsilon_\alpha + P_\alpha) \
u_{\alpha}^{\mu} \ u_{\alpha}^{\nu} -g^{\mu\nu} P_\alpha .
   \label{eq11}
   \end{eqnarray}
 For the fireball, however, only
the thermalized part of the energy--momentum tensor is described
by this hydrodynamic form
   \begin{eqnarray}
T^{\scr{(eq)}\mu\nu}_{\scr f}=
(\varepsilon_{\scr f} + P_{\scr f}) \ u_{\scr f}^{\mu} \ u_{\scr f}^{\nu}
-g^{\mu\nu} P_{\scr f}.
   \label{eq12}
   \end{eqnarray}
Its evolution is defined by the Euler equation with a retarded
source term
   \begin{eqnarray}
   \partial_{\mu} T^{\scr{(eq)}\mu\nu}_{\scr f} (x) &=&
\int d^4 x' \delta^4 \left(\vphantom{I^I_I}
x - x' - U_F (x')\tau\right)
 \left[F_{\scr p}^\nu (x') + F_{\scr t}^\nu (x')\right] \nonumber
 \\
&-& F_{\scr{fp}}^\nu (x) - F_{\scr{ft}}^\nu (x),
   \label{eq13}
   \end{eqnarray}
where $\tau$ is the formation time, and
   \begin{eqnarray}
   \label{eq14}
U^\nu_F (x')=
\frac{F_{\scr p}^\nu(x')+F_{\scr t}^\nu(x')}%
{|F_{\scr p}(x')+F_{\scr t}(x')|}
   \end{eqnarray}
is a free-streaming 4-velocity of the produced fireball matter. In
fact, this is the velocity at the moment of production of the
fireball matter. According to Eq.~(\ref{eq13}), the energy and
momentum of this matter appear as a source in the Euler equation
only later, at the time $U_F^0\tau$ after production, and in
different space point ${\bf x}' - {\bf U}_F (x')\tau$, as compared
to the production point ${\bf x}'$.

The residual part of $T^{\mu\nu}_{\scr f}$ (the free-streaming
one) is defined as
   \begin{eqnarray}
T^{\scr{(fs)}\mu\nu}_{\scr f}=
T^{\mu\nu}_{\scr f}-T^{\scr{(eq)}\mu\nu}_{\scr f}.
   \label{eq15}
   \end{eqnarray}
The equation for $T^{\scr{(eq)}\mu\nu}_{\scr f}$ can be easily
obtained by taking the difference between Eqs.~(\ref{eq8f}) and
(\ref{eq13}). If all  the fireball matter turns out to be formed
before freeze-out (which is the case, in fact), then this equation
is not needed. Thus, the 3-fluid model introduced here contains
both the original  2-fluid model with pion radiation
\cite{MRS88,MRS91,INNTS} and the (2+1)-fluid model
\cite{Kat93,Brac97} as limiting cases for $\tau \rightarrow
\infty$ and $\tau=0$, respectively.

The nucleon--nucleon cross sections at high energies are strongly
forward--backward peaked. Since the involved 4-momentum transfer
is small, the Boltzmann collision term  can be essentially
simplified and in this case the friction forces, $F_{\scr p}^\nu$
and $F_{\scr t}^\nu$, are estimated  as
 \begin{eqnarray}
 F_{\alpha}^\nu=\rho_{\scr p} \rho_{\scr t}
\left[\left(u_{\alpha}^{\nu}-u_{\bar{\alpha}}^{\nu}\right)D_P+
\left(u_{\scr p}^{\nu}+u_{\scr t}^{\nu}\right)D_E\right],
\label{eq16}
\end{eqnarray}
$\alpha=$p and t, $\bar{\mbox{p}}=$t and  $\bar{\mbox{t}}=$p. Here,
$\rho_\alpha$ denotes the scalar densities of the p- and t-fluids,
 \begin{eqnarray}
D_{P/E} = m_N \ V_{\scr{rel}}^{\scr{pt}} \
\sigma_{P/E} (s_{\scr{pt}}),
\label{eq17}
\end{eqnarray}
where $m_N$ is the nucleon mass, $s_{\scr{pt}}=m_N^2 \left(u_{\scr
p}^{\nu}+u_{\scr t}^{\nu}\right)^2$ is the mean invariant energy
squared of two colliding nucleons from the p- and t-fluids,
$V_{\scr{rel}}^{\scr{pt}}=
[s_{\scr{pt}}(s_{\scr{pt}}-4m_N^2)]^{1/2}/2m_N^2$ is their mean
relative velocity, and $\sigma_{P/E}(s_{\scr{pt}})$ are determined
in terms of nucleon-nucleon cross sections integrated with certain
weights (see \cite{MRS88,MRS91,Sat90} for details).

Eqs.~(\ref{eq8})--(\ref{eq8t}) and (\ref{eq13}), supplemented by a
certain equation of state (EoS) and expressions for friction
forces $F^\nu$, form a full set of equations of the relativistic
3-fluid hydrodynamic model. To make this set closed, we still need
to define the friction of the fireball fluid with the p- and
t-fluids, $F_{\scr{fp}}^\nu$ and $F_{\scr{ft}}^\nu$, in terms of
hydrodynamic quantities and some cross sections.

To estimate the scale of the friction force between the fireball
and baryon-rich fluids, similar to that done above for baryon-rich
fluids, we consider a simplified system, where all baryon-rich
fluids consist only of nucleons, as the most abundant component of
these fluids, and the fireball fluid contains only pions.  At
incident energies from 10 (AGS) to 200 A$\cdot$GeV (SPS) the
relative nucleon-pion energies prove to be in the resonance range
dominated by the $\Delta$-resonance. The resonance-dominated
interaction implies that the essential process is absorption of a
fireball pion by a p- or t-fluid nucleon with formation of an
$R$-resonance (most probably $\Delta$).  As a consequence,  only
the loss term contributes to the kinetic equation for the fireball
fluid. After momentum integrating this collision term weighted
with the 4-momentum $p^\nu$, we  get~\cite{RIN03}
\begin{eqnarray}
\hspace*{-7mm}
F_{\scr{f}\alpha}^\nu (x)
=
 D_{\scr{f}\alpha} \ \frac{T^{\scr{(eq)}0\nu}_{\scr f}}{u_{\scr
 f}^0} \ \rho_{\alpha}~,
\label{eq19}
\end{eqnarray}
where transport coefficients take the form
\begin{eqnarray}
D_{\scr{f}\alpha} = W^{N\pi\to R}(s_{{\scr f}\alpha})/(m_N m_\pi)
=V_{\scr{rel}}^{{\scr f}\alpha} \ \sigma_{\scr{tot}}^{N\pi\to
R}(s_{{\scr f}\alpha}).
\end{eqnarray}
Here, $V_{\scr{rel}}^{{\scr f}\alpha}=[(s_{{\scr
f}\alpha}-m_N^2-m_\pi^2)^2 -4m_N^2m_\pi^2]^{1/2}/(2m_N m_\pi)$
denotes the mean invariant relative velocity between the fireball
and the $\alpha$-fluids, $s_{{\scr f}\alpha} = (m_\pi u_{\scr
f}+m_N u_{\alpha})^2$, and $\sigma_{\scr{tot}}^{N\pi\to R}(s)$ is
the parameterization of experimental pion--nucleon cross-sections.
Thus, we have expressed the friction $F_{\scr{f}\alpha}^\nu$ in
terms of the fireball-fluid energy-momentum density
$T^{0\nu}_{\scr f}$, the scalar density $\rho_{\alpha}$ of the
$\alpha$ fluid, and a transport coefficient $D_{\scr{f}\alpha}$.
Note that this friction is zero until the fireball pions are
formed, since $T^{\scr{(eq)}0\nu}_{\scr f}=0$ during the formation
time $\tau$.

In fact, the above treatment is an estimate of the friction terms
rather than their strict derivation. As it is seen from
Eq.~(\ref{eq17}) for the excited matter of baryon-rich fluids, a
great number of hadrons and/or deconfined quarks and gluons may
contribute into this friction. Furthermore, these quantities may
be modified by in-medium effects. In this respect, $D_{P/E}$ and
$D_{\scr{f}\alpha}$ should be understood as quantities that give a
scale of this interaction.

\section{Simulations of Nucleus--Nucleus Collisions}

The relativistic 3D code for the above described 3-fluid model was
constructed by means of modifying the existing 2-fluid 3D code of
Refs. \cite{MRS88,MRS91,INNTS}. In actual calculations we used the
mixed-phase EoS developed in \cite{NST98,TNS98}. This
phenomenological EoS takes into account a possible deconfinement
phase transition of nuclear matter. The underlying assumption of
this EoS is that  unbound quarks and gluons may coexist with
hadrons in the nuclear environment. In accordance with lattice QCD
data, the statistical mixed-phase model describes the first-order
deconfinement phase transition for pure gluon matter and crossover
for that with quarks \cite{NST98,TNS98}. For details concerning
the used EoS's, please, refer to \cite{TNFNR03}.

In Fig. 1 global dynamics of heavy-ion collisions is illustrated
by the energy-density evolution of the baryon-rich  fluids
($\varepsilon_{\scr{b}}=\varepsilon_{\scr{p}}+\varepsilon_{\scr{t}}$)
in the reaction plane of the Pb+Pb collision at $E_{\scr{lab}}=$
158 A$\cdot$GeV. Different stages of interaction at relativistic
energies are clearly seen in this example: Two Lorentz-contracted
nuclei (note the different scales along the $x$- and $z$-axes in
Fig.1) start to interpenetrate through each other, reach a maximal
energy density by the time $\sim 1.1$ fm/c and then expand
predominantly in longitudinal direction forming a "sausage-like"
freeze-out system. At this and lower incident energies the
baryon-rich dynamics is not really disturbed by the fireball fluid
and hence the cases $\tau=$ 0 and 1 fm/c turned to be
indistinguishable in terms of $\varepsilon_{\scr{b}}$.

Time--evolution  of $\varepsilon_{\scr{b}}$ in Fig.1  is
calculated for the mixed phase model.  Topologically results for
EoS of pure hadronic and that of two-phase models look very
similar. Due to essential softening the equation of state near the
deconfinement phase transition, in the last case the system
evolves noticeably slower what may have observable
consequences~\cite{NST98,TNS98}.
\begin{figure}
\begin{center}
\includegraphics[height=17cm,clip]{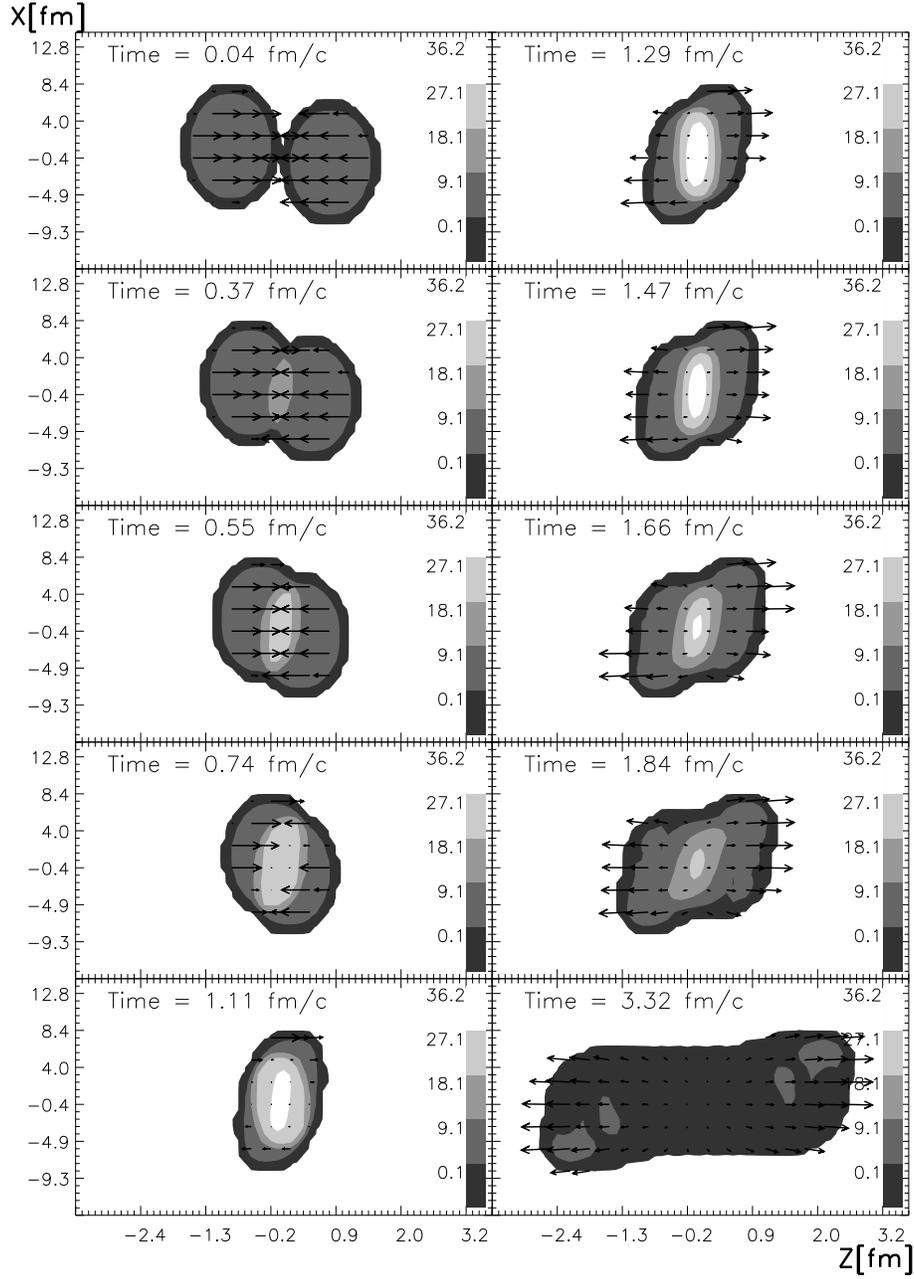}
\caption[C1]{Time evolution of the  energy density,
$\varepsilon_{\scr{b}}=\varepsilon_{\scr{p}}+\varepsilon_{\scr{t}}$,
 for  the baryon-rich fluids in the
reaction plane ($xz$ plane) for the Pb+Pb collision
($E_{\scr{lab}}=$ 158 A$\cdot$GeV) at impact parameter $b=$ 2 fm.
Shades of gray represent different levels of
$\varepsilon_{\scr{b}}$ as indicated at the right side of each
panel. Numbers at this palette show the $\varepsilon_{\scr{b}}$
values (in GeV/fm$^3$) at which the shades change. Arrows indicate
the hydrodynamic velocities of the fluids. } \label{fig1}
\end{center}
\end{figure}
\begin{figure}
\begin{center}
\includegraphics[height=11cm,angle=-90,clip]{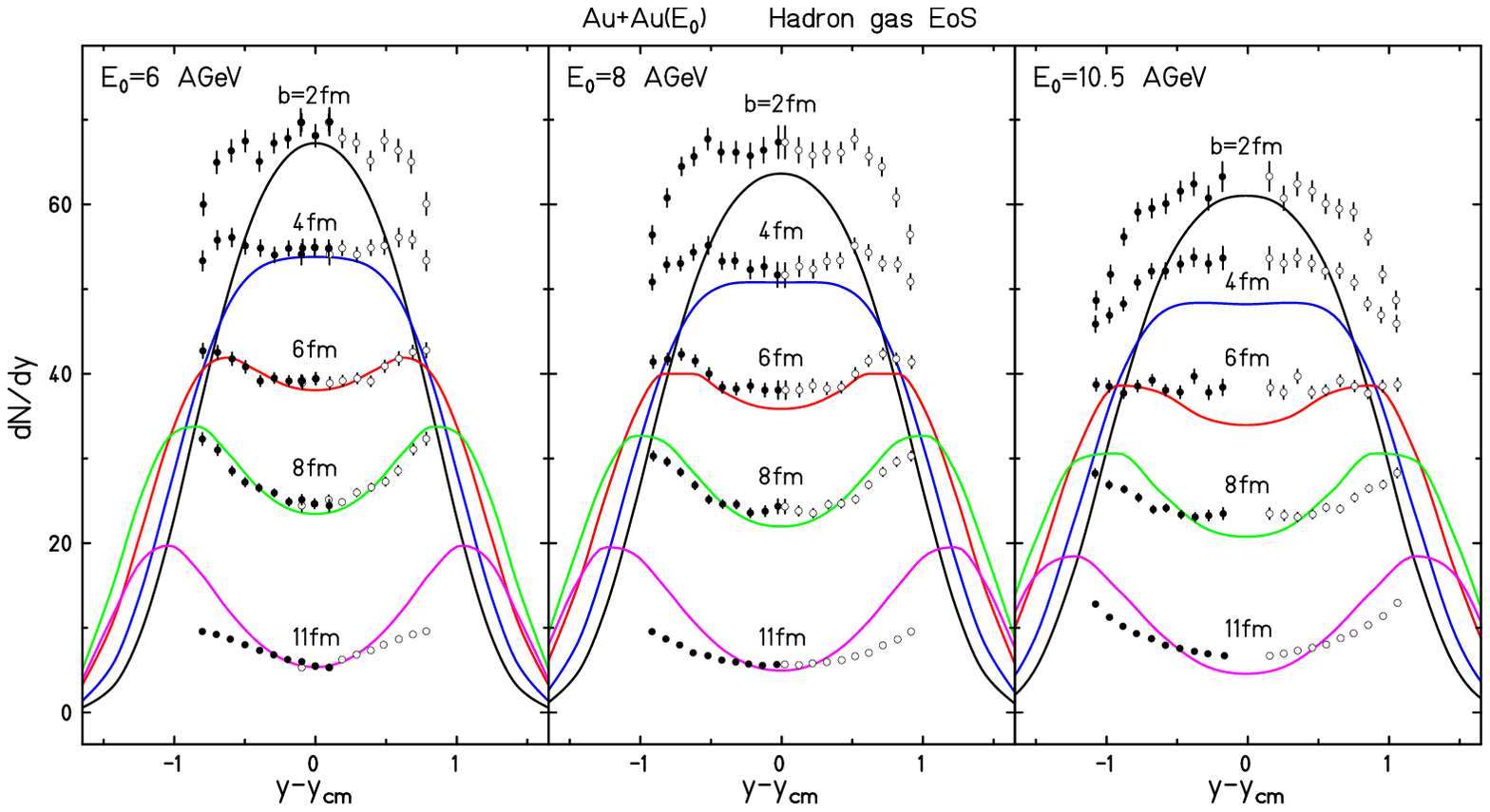}
\includegraphics[height=11cm,angle=-90,clip]{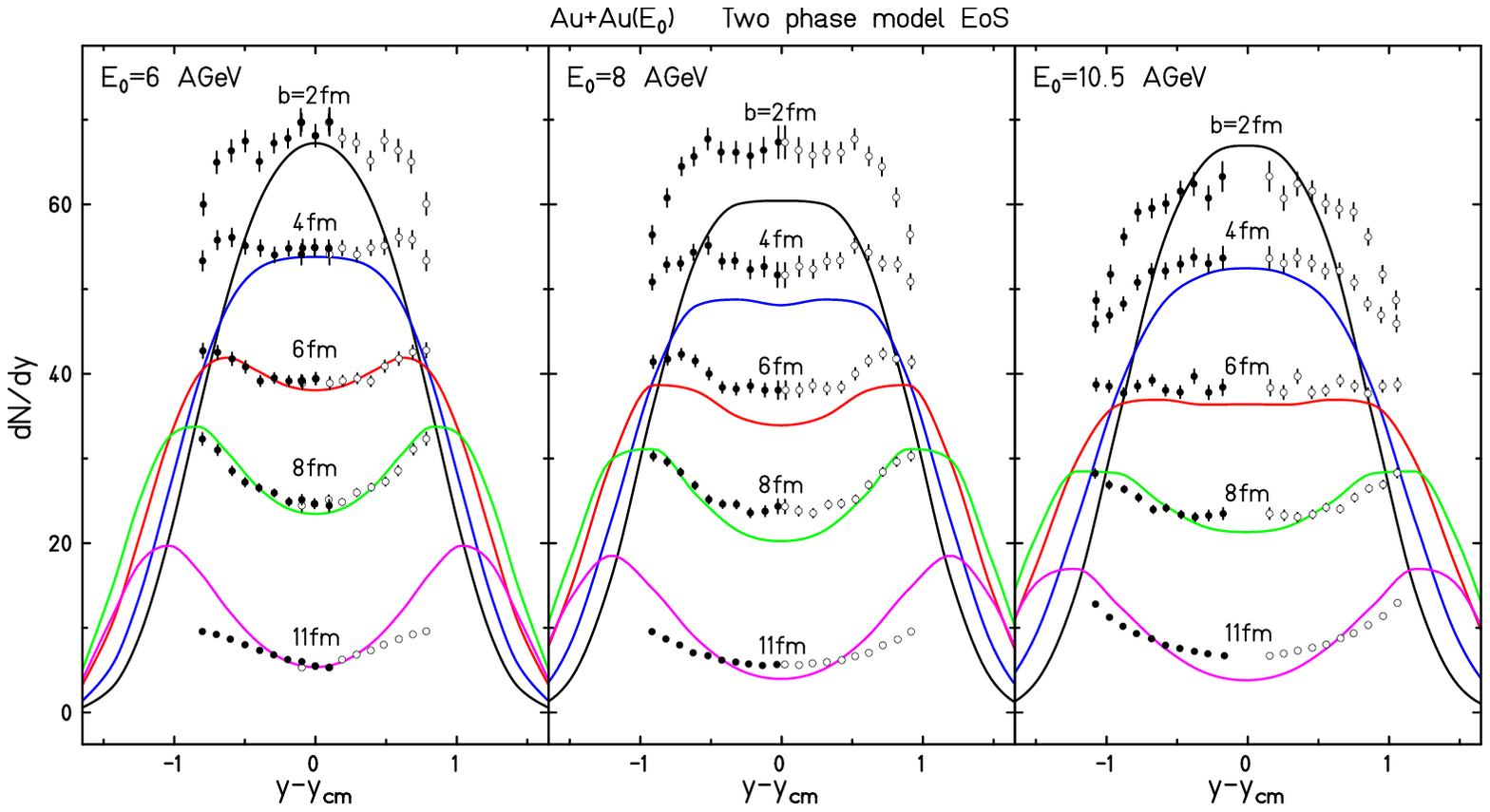}
\includegraphics[height=11cm,angle=-90,clip]{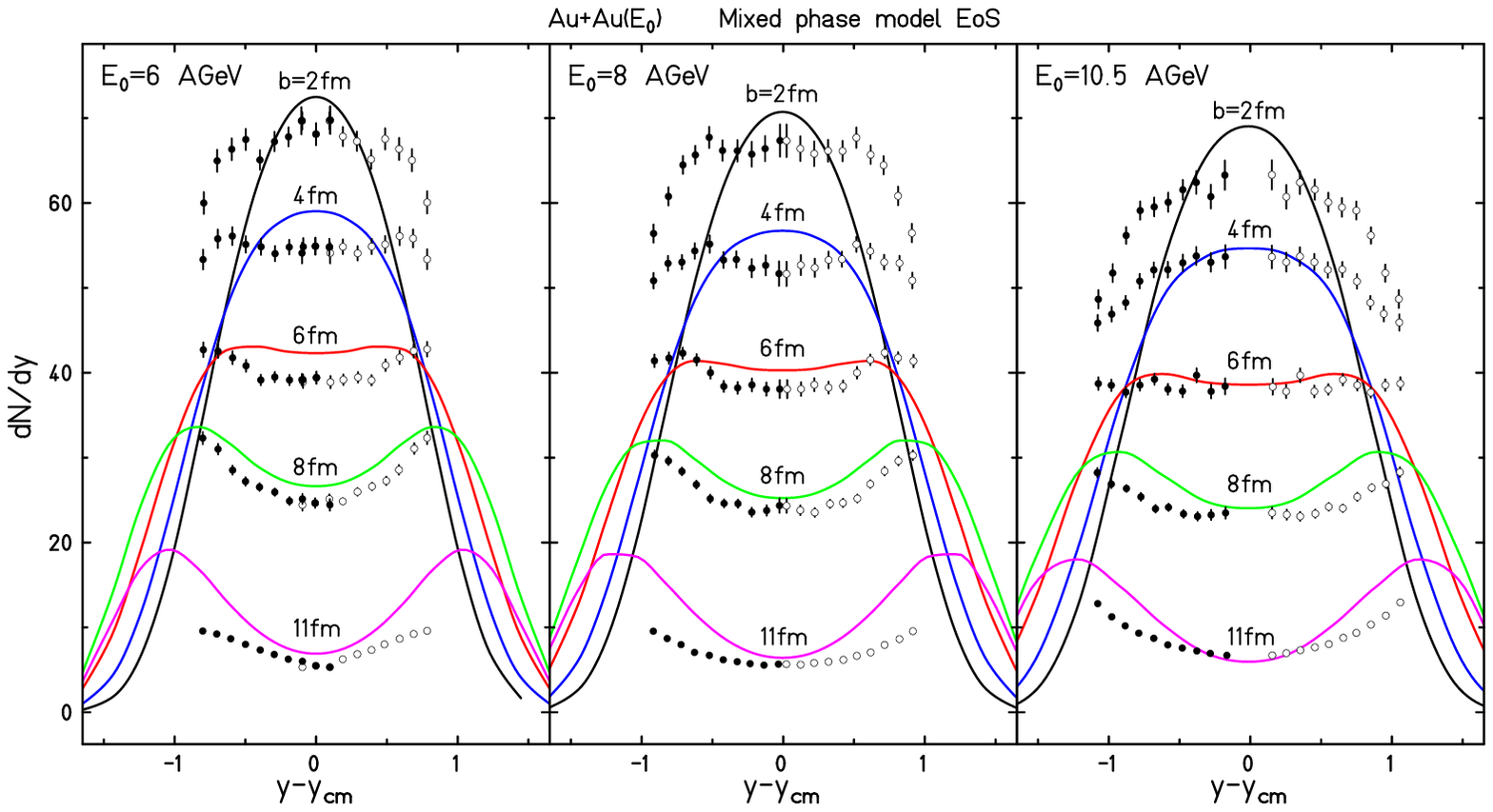}
\caption[C2]{Proton rapidity spectra from $Au+Au$ collisions at
three bombarding energies and different impact parameters
$b$(given in fm) Three panels correspond to different equation of
state. Experimental points are from~\cite{E917} } \label{fig2}
\end{center}
\end{figure}

The energy released in the fireball fluids is  of an order of
magnitude smaller than that stored in baryon--rich fluids and
depends on the formation time. At realistic values of the
formation times, $\tau\sim$ 1 fm/c, the effect of the interaction
is substantially reduced. It happens because the fireball fluid
starts to interact only near the end of the interpenetration
stage. As a result, by the end of the collision process it looses
only 10\% of its available energy at $E_{\scr{lab}}=$ 158
A$\cdot$GeV and 30\%, at $E_{\scr{lab}}=$ 10.5 A$\cdot$GeV.
Certainly, this effect should be observable in mesonic quantities,
in particular, in such fine observables as directed and elliptic
flows. The global baryonic quantities stay practically  unchanged
at finite $\tau$~\cite{RIN03}.

To calculate observables, hydrodynamic calculations should be
stopped at some freeze-out point. In our model it is assumed
 that a fluid element decouples from the hydrodynamic regime,
when its energy density $\varepsilon$ and densities in the eight
surrounding cells become smaller than a fixed  value
$\varepsilon_{fr}$. A value $\varepsilon_{fr} = 0.15 \ GeV/fm^3$
was used for this local freeze-out density which corresponds to
the actual energy density of the freeze-out fluid element of $\sim
0.12 \ GeV/fm^3$. To proceed to observable free hadron gas, the
shock-like freeze-out~\cite{Bug96} is assumed, conserving energy
and baryon charge.

Proton rapidity spectra calculated for $Au+Au$ collisions are
presented in Fig.2 for $E_{beam}=$6, 8 and 10.5 $A\cdot GeV$. One
should note that hydrodynamics does not make difference between
bounded into fragments and free nucleons.  In the results
presented the contribution from light fragments (d, t, $^3$He and
$^4$He) has been  subtracted using a simple coalescence
model~\cite{RIPH94}.
\begin{figure}[h]
\begin{center}
\includegraphics[height=10cm,angle=-90,clip]{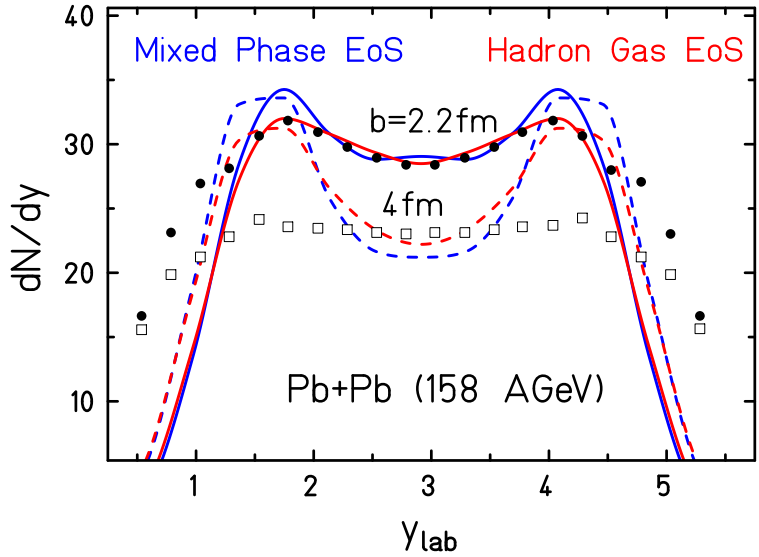}
\vspace*{-5mm} \caption[C3]{Rapidity spectra of protons from
$Pb+Pb$ collisions for the mixed phase and hadron gas EoS. Solid
and dashed lines are calculated for impact parameters 2 and 4 fm,
respectively. Experimental points are from~\cite{NA49_99}.}
\label{fig3}
\end{center}
\end{figure}
This procedure allows one to reproduce reasonably evolution of the
spectra shape  with changing the impact parameter $b$ for all the
energies considered. Some discrepancy observed for peripheral
collisions near the target and projectile rapidity are mainly due
to a not-subtacted contribution of heavier fragments. As seen, the
rapidity spectra are only slightly sensitive to the EoS used. The
same conclusions can be drawn from Fig.3 for higher CERN SPS
energies. One should notice that  a general agreement here for
central collisions is much better than that reached in any other
2-fluid hydrodynamic model. This originates mainly from larger
energy-momentum transfer in (\ref{eq16}) rather than from the
account for the third fireball fluid.

\section{Conclusions}
In this paper we have developed a 3-fluid model for simulating
heavy-ion collisions in the  range of incident energies between
few and about 200 A$\cdot$GeV. In addition to two baryon-rich
fluids, which constitute the 2-fluid  model, a delayed evolution
of the produced baryon-free (fireball) fluid is incorporated. This
delay is governed by a formation time, during which the fireball
fluid neither thermalizes nor interacts with the baryon-rich
fluids. After the formation, it thermalizes and comes into
interaction with the baryon-rich fluids. This interaction is
estimated from elementary pion-nucleon cross-sections.
Implementation of different EoS, including those with the
deconfinement phase transition,  may open great opportunities for
analysis of collective effects in relativistic heavy-ion
collisions. Unfortunately, in spite of reasonable reproduction of
observable proton rapidity spectra in the wide range of bombarding
energies and centrality parameters we are unable to favor any of
considered EoS. Analysis of more delicate characteristics is
needed. This work is in progress now.

\vspace*{5mm} {\bf Acknowledgements}  This work was supported in
part by the Deutsche Forschungsgemeinschaft (DFG project 436 RUS
113/558/0-2), the Russian Foundation for Basic Research (RFBR
grant 03-02-04008) and Russian Minpromnauki (grant
NS-1885.2003.2).


\begin{thebibliography}{99}\itemsep=0pt

\bibitem{RIN03}V.N.~Russkikh, Yu.B.~Ivanov, E.G.~Nikinov,
W.~N\"{o}renberg, and V.D.~Toneev, Yad. Fiz. \textbf{67},  (2004)
(to be published) [nucl-th/0302029].
%
\bibitem{Bl57}D.I.~Blokhintsev, JETP \textbf{32},  350
(1957) [in Russian].
%
\bibitem{MRS88}  I.N.~Mishustin, V.N.~Russkikh, and
L.M.~Satarov, Yad. Fiz. {\bf 48}, 711 (1988) [Sov. J. Nucl. Phys.
{\bf 48},  454 (1988)]; Nucl. Phys. {\bf A494}, 595 (1989).
   Nucl. Phys. {\bf A446}, 727 (1985).
%
\bibitem{I87}
Yu.B.~Ivanov,
     Nucl. Phys. {\bf A474},  669 (1987).
%
\bibitem{MRS91}   I.N.~Mishustin, V.N.~Russkikh, and
L.M.~Satarov, Yad. Fiz. {\bf 54}, 429 (1991) [Sov. J. Nucl. Phys.
{\bf 54}, 260 (1991)];  I.N.~Mishustin, V.N.~Russkikh, and
L.M.~Satarov, in {\em Relativistic heavy ion physics},
 L.P.~Csernai and  D.D.~Strottman (eds), (World Scientific, 1991) p.179.
%
\bibitem{INNTS}
      Yu.B.~Ivanov, E.G.~Nikonov, W.~N\"orenberg, V.D.~Toneev and
      A.A.~Shanenko,  Heavy Ion Phys. {\bf 15}, 127 (2002)
      [nucl-th/0011004].

\bibitem{Kat93}
U.~Katscher, D.H.~Rischke, J.A.~Maruhn, W.~Greiner, I.N.~Mishustin,
and L.M.~Satarov, Z. Phys. {\bf A346}, 209 (1993);
%
U.~Katscher, J.A.~Maruhn, W.~Greiner, and  I.N.~Mishustin, Z. Phys.
{\bf A346}, 251 (1993);
%
A.~Dumitru, U.~Katscher, J.A.~Maruhn, H.~St\"ocker, W.~Greiner, and
D.H.~Rischke, Phys. Rev. {\bf C51}, 2166 (1995) [hep-ph/9411358];
%
Z. Phys. {\bf A353}, 187 (1995) [hep-ph/9503347].
%
\bibitem{Brac97}
By J.~Brachmann, A.~Dumitru, J.A.~Maruhn, H.~St\"ocker, W.~Greiner, and
D.H.~Rischke, Nucl. Phys. {\bf A619}, 391 (1997) [nucl-th/9703032];
%
A.~Dumitru, J.~Brachmann, M.~Bleicher, J.A.~Maruhn, H.~St\"ocker, and
W.~Greiner, Heavy Ion Phys. {\bf 5}, 357 (1997) [nucl-th/9705056];
%
M.~Reiter, A.~Dumitru, J.~Brachmann, J.A.~Maruhn, H.~St\"ocker, and
W.~Greiner, Nucl. Phys. {\bf A643}, 99 (1998) [nucl-th/9806010];
%
M.~Bleicher, M.~Reiter, A.~Dumitru, J.~Brachmann, C.~Spieles,
S.A.~Bass, H.~St\"ocker,  and W.~Greiner, Phys. Rev. {\bf C59},
1844 (1999) [hep-ph/9811459].
%
%
%
\bibitem{Sat90} L.M.~Satarov, Yad. Fiz. {\bf 52}, 412 (1990)
[Sov. J. Nucl. Phys. {\bf 52}, 264 (1990)].
%
\bibitem{NST98} E.G.~Nikonov, A.A.~Shanenko, and V.D.~Toneev,
Heavy Ion Phys. {\bf 8}, 89 (1998) [nucl-th/9802018]; Yad. Fiz.
{\bf 62}, 1301 (1999) [Physics of Atomic Nuclei {\bf 62},  1226
(1999)].
%
\bibitem{TNS98}  V.D.~Toneev, E.G.~Nikonov, and A.A.~Shanenko,
in {\em Nuclear Matter in Different Phases and Transitions},
eds. J.-P. Blaizot, X.~Campi, and M.~Ploszajczak (Kluwer Academic
Publishers, 1999) p.309.
%
\bibitem{TNFNR03} V.D.~Toneev, E.G.~Nikonov,  B.~Friman, W.~N\"orenberg,
K.~Redlich,   hep-ph/0308088.
%
\bibitem{E917}  E917 Collaboration,  Phys. Rev. Lett. \textbf{86},  1970
(2001) [nucl-ex/0003007].
%
\bibitem{Bug96}
K.A.~Bugaev, Nucl.Phys. \textbf {A606}, 59 (1996).
%
\bibitem{RIPH94}V.N.~Russkikh, Yu.B.~Ivanov, Yu.E.~Pokrovsky, and P.A.~Henning,
Nucl. Phys. \textbf{A572}, 749 (1994).
%
\bibitem{NA49_99} G.E.~Cooper {\it et al}, Nucl. Phys.,
\textbf{A661} 362c (1999).

\end{thebibliography}
 \end{document}